To the Production office of EPJ B
                                        your Ref. Ms B9073

Please, find enclosed the following revtex version of the manuscript
``Optical properties of an interacting large polaron gas'' by 
V. Cataudella, G. De Filippis and G. Iadonisi accepted for publication.
The archive contains also three figures in ps format. 
Yours sincerely,           V. Cataudella

\documentstyle[12pt,aps,prb]{revtex}
\begin{document} 
  
\title{Optical properties of an interacting large polaron gas}
\author{V.~Cataudella$^+$, G.~De~Filippis$^*$,
G.~Iadonisi$^+$}
\address{$^+$INFM and Dipartimento di Scienze Fisiche, 
Universit\`a di Napoli I-80125 Napoli, Italy} 
\address {$^*$INFM and Dipartimento di Scienze Fisiche, 
Universit\`a di Salerno I-84081 Baronissi (Salerno), Italy} 
\date {\today} 
\maketitle 
\begin {abstract} 
The normal state conductivity, $\sigma(\omega)$, of a 
system of interacting large polarons is calculated within the Random
Phase approximation and some numerical results are presented. 
The behaviour of the 
optical absorption as a function of the charge carrier density and of the 
temperature is analyzed for different values of the electron-phonon 
coupling constant. 
It is shown that $\sigma(\omega)$ exhibits features similar to those observed 
in the infrared spectra of the cuprates. 
\end {abstract}

\begin{tabular}{ll}
71.38:   & polarons and electron-phonon interaction\\
\end{tabular} 
\pacs{PACS: 71.38 (Polarons)  } 

\newpage
\section{Introduction.}
The infrared absorption of large polarons is a well studied 
problem.\cite{Devreese_review} However, a large part of the studies has been
 focused mainly on the very low density regime (non interacting 
 polaron limit) which is suitable for not heavily doped polar
 semiconductors and ionic insulators. In  this regime the most accurate
 approach  has been discussed by Devreese et al.\cite{Devreese_absorption}
 starting from the expression of the impedance function derived by
 Feynmann.\cite{21} More recently the interest for this problem
 has been renewed in connection with the infrared absorption of
cuprates in the normal phase.\cite{review calvani,emin} In this materials the
interaction among polarons can be important.
 It is, therefore,
of interest to consider the effect of the Coulomb long range interaction
on the large polaron absorption. To our knowledge this effect has not been
yet considered within the Devreese-Feynman approach to the polaron
absorption.

The aim of this paper is to calculate the conductivity $\sigma(\omega)$
of a large polaron gas within the random phase approximation (RPA)
by using, for the lowest order polarization insertion, the propagator of
the Feynman polaron model.\cite {20,Schultz} We find that it is
possible to identify  in  $\sigma(\omega)$ different 
contributions with features common to some structures of the 
infrared spectra of cuprates. In particular 
we observe a significant displacement of spectral weight from higher
to lower frequencies when the polaron density increases.

\section{The model.} 
We consider a system made of band electrons in the effective mass approximation
interacting with non 
dispersive LO phonons and repelling each 
other through the Coulomb potential screened by 
the background high frequency dielectric constant  
$\epsilon_{\infty}$\cite{16}. 
The electron-phonon (e-ph) interaction is assumed to be: 
\begin {equation}
H_{e-ph}=\sum_{\stackrel{\vec{p}\vec{q}}{\sigma}}
M_qc^{\dagger}_{\vec{p}+\vec{q}\sigma}c_{\vec{p}\sigma}\left(a_{\vec{q}}+
a^{\dagger}_{-\vec{q}} \right)                                                                      
\label{1r} 
\end {equation}
where
\begin {equation} 
M_{q}=-i\hbar\omega_{0}\frac{R^{1 / 2}_{p}}{q}\left(\frac{4\pi\alpha}{V} 
\right)^{1/2}  
\label{2r} 
\end {equation} 
is the Fr\"ohlich e-ph matrix element\cite{17}. 

In the above expressions  
$c_{\vec{p}\sigma}$ ($c^{\dagger}_{\vec{p}\sigma}$) and 
$a_{\vec{q}}$ ($a^{\dagger}_{\vec{q}}$) indicate, respectively, 
the annihilation (creation) operators for electrons and phonons, 
$V$ is the system's volume, $\alpha$ is the Fr\"ohlich e-ph coupling 
constant, $R_{p}=\left(\hbar/\left(2m\omega_{0}\right)\right)^{1/2}$ is the 
polaron radius and $\omega_0$ is the LO phonon frequency. 

The normal state conductivity of this system, $\sigma(\omega)$, is related to 
the retarded form of the current-current correlation function $\Pi(\omega)$ by 
the Kubo formula\cite{18}: 
\begin {equation} 
\sigma(\omega)=\frac{i}{\omega}\left(\frac{ne^2}{m}+\Pi(\omega)\right) 
\label{3r} 
\end {equation} 
where $n$ is the density of charge carriers. In order to estimate the
conductivity we follow the approach suggested by Feynman\cite{21} and
adopted with success by Devreese et al.\cite{Devreese_absorption}
for the optical absorption of non interacting large polarons.
Within this approach the real part of the conductivity is given by:

\begin {equation} 
Re\left[\sigma(\omega)\right]=-G(\omega)\frac{Im\Sigma(\omega)}{\left(\omega-
Re\Sigma(\omega)\right)^2+\left(Im\Sigma(\omega)\right)^2}
\label{10r} 
\end {equation} 
where $G(\omega)=ne^2/m$ and $\Sigma(\omega)=\Pi^0(\omega)\omega/G(\omega)$
(for the definition of $\Pi^0(\omega)$ see below).
 A justification of eq.(\ref{10r}) has been given by Feynman\cite{21}, while 
a more general derivation can be obtained by making use of the 
Mori-Zwanzig\cite{22} projection operator technique, as shown by Devreese 
et al.\cite{23} and Maldague\cite{maldague}. 
It is worth to note that the sum rule for the real part of the conductivity: 
\begin {equation} 
\int_{0}^{\infty} d\omega Re\left[\sigma(\omega)\right]=\frac{\pi ne^2}{2m} 
\label{11r}
\end {equation} 
is satisfied by using eq.(\ref{10r}).

To evaluate $\Pi^0(\omega)$ we use the set of diagrams shown in Fig.1, i.e., 
we 
consider a R.P.A.-like approximation for polarons interacting each other 
through the Coulomb potential screened by $\epsilon_{\infty}$ 
($v_q^{\infty}$). 
Following Mahan\cite{19}, we write: 
\begin {equation} 
\Pi^0(\omega)=\frac{-e^2}{m^2\omega^2}\int \frac {d^3q}
{\left(2\pi\right)^3} q_{\mu}^2\left[N(\vec{q},\omega)-N(\vec{q},0)\right]  
\label{4r} 
\end {equation} 
where 
\begin {equation} 
N(\vec{q},\omega)=\int_{-\infty}^{\infty} \frac{du}{\pi} 
\int_{-\infty}^{\infty} \frac{ds}{\pi} \frac {1}
{v_q^{\infty}}Im\left[\frac{1}{\epsilon(\vec{q},u)}\right]Im\left[
\bar{M_q}^2D(\vec{q},s)
\right]\left(\frac{n_{\scriptscriptstyle{B}}(s)-n_{\scriptscriptstyle{B}}(u)}
{u-s+\omega+i\delta}\right). 
\label{5r} 
\end {equation} 
In the eqs.(\ref{4r}) and (\ref{5r}) $\mu=x,y$ or $z$, 
$\epsilon(\vec{q},\omega)$ is the dielectric function, $\bar{M_{\vec{q}}}$ 
is the 
renormalized e-ph matrix element, $D(\vec{q},\omega)$ is the 
phonon Green function and $n_{\scriptscriptstyle{B}}(\omega)$ is the boson 
occupation number. 
In our model the dielectric function of the system is assumed to be (see 
Fig.1): 
\begin {equation} 
\epsilon(\vec{q},\omega)=1-v_q^{\infty}P(\vec{q},\omega) 
\label{6r} 
\end {equation} 
where $P(\vec{q},\omega)$, the lowest order polarization insertion, can be 
expressed in 
terms of the polaron spectral weight function $A(\vec{q},\omega)$: 
\begin {equation} 
P(\vec{q},\omega)=\frac{2}{\hbar^2} \int \frac {d^3p}{\left(2\pi\right)^3}
\int_{-\infty}^{\infty} \frac {du}{2\pi} \int_{-\infty}^{\infty} 
\frac {ds}{2\pi} A(\vec{p},s)A(\vec{p}+\vec{q},u)\left(\frac
{n_{\scriptscriptstyle{F}}(s)-n_{\scriptscriptstyle{F}}(u)}
{\omega+s-u+i\delta}\right). 
\label{7r} 
\end {equation} 
Here $n_{\scriptscriptstyle{F}}(\omega)$ is the fermion 
occupation number.   
For simplicity we shall use in the Eq.(\ref{5r}) the unperturbed phonon Green 
function $D^0(\omega)=2\omega_0/\left(\omega^2-\omega_0^2+i\delta\right)$ 
and we shall ignore the frequency dependence of the dielectric function in the 
expression of the renormalized e-ph matrix element assuming: 
\begin {equation} 
\bar{M_{\vec{q}}}=\frac{M_q}{\epsilon(\vec{q},0)}. 
\label{8r} 
\end {equation} 
Until this point we have not specified the polaron spectral weight function 
that appears in the Eq.(\ref{7r}).
It is well known that, from a theoretical point of view, the problem of an 
electron in a polar crystal which interacts with the 
longitudinal optical modes of lattice vibrations has not been solved 
exactly. However, it is universally recognized that of all published theories 
that of Feynman\cite{20}, which uses a variational method based on path 
integrals, gives the best available results in the entire range of the 
coupling constant $\alpha$. For this reason we choose the 
spectral weight function of the Feynman polaron model\cite{20,Schultz}: 
 
\begin {equation} 
A(q,\omega)=\sum_{l=-\infty}^{\infty}\delta\left[\omega-\left(\frac{\hbar^2q^2}
{2M_T}+lv\right)\right]exp{\left[\frac{l\beta v}{2}\right]}exp{
\left[-\frac{q^2R}{M_T}
\left(2n_{\scriptscriptstyle{B}}(v)+1\right)\right]} 2 \pi I_l(z) 
\label{9r} 
\end {equation} 
where $R=\left(\frac{v^2}{w^2}-1\right)/v$,
$M_T=v^2/w^2$, 
$z=\frac{2q^2R}{M_T}\left[n_{\scriptscriptstyle{B}}(v)\left(
n_{\scriptscriptstyle{B}}(v)+1\right)\right]^{1/2}$ 
and $I_l$ are the Bessel functions of complex argument.   
The dimensionless parameters $v$ and $w$ are related to the mass and the 
elastic constant of the model\cite{20,Schultz}, in which the electron is 
coupled via a harmonic force to 
a fictitious second particle simulating the phonon degrees of freedom. 
The values of $v$ and $w$ can be obtained by the variational approach 
described by Feynman\cite{20} and Schultz\cite{Schultz}. In particular, 
at $T=0$, $A(q,\omega)$ takes the form: 

\begin {equation} 
A(q,\omega)=2 \pi\sum_{l=0}^{\infty}\delta\left[\omega-\left(\frac{\hbar^2q^2}
{2M_T}+lv\right)\right]e^{-\frac{q^2R}{M_T}} 
\left(\frac{q^2R}{M_T}\right)^{l}/l!~.   
\label{50r} 
\end {equation} 
The spectral weight follows a Poissonian distribution and it is maximum for 
an excitation involving a number $l$ of phonons of order of 
$l\sim\frac{q^2R}{M_T}$. We note that, in all the numerical results which 
will be shown in this paper, the terms with $-6\le l \le6$ in the 
Eq.(\ref{9r}) 
and the first six term in the Eq.(\ref{50r}) are enough to a good convergence 
of the real and imaginary parts of the correlation function $\Pi^0(\omega)$ in 
the frequency range of interest ($\omega\le 10\omega_0$).

We want to note that the approach proposed in this paper restores, 
when $n \rightarrow 0$ ,
the well known results of the optical absorption of a single polaron\cite{23}  
and allows to introduce within the R.P.A. approximation the effects of the 
polaron-polaron interaction. This approach is expected to be valid 
when the formation of bipolaron states is not favored 
($\eta=\epsilon_{\infty}/\epsilon_0>.01$ or 
$\eta<0.01$ and $\alpha<6-7$)\cite{iadonisi}. 
In fact, in this situation   
the overlap between the wells of two particles can be neglected and the system 
is well approximated by an interacting large polaron gas. Of course the
proposed approach is valid in the density range where we can exclude 
Wigner-like localization. As discussed by Quemerais et al.\cite{quemerais}
for coupling constant not larger than $\alpha=6$ and density $n
\geq 10^{18}cm^{-3}$ our approach is justified.  

\section{ The results} 

In Fig.2a is reported the optical 
absorption per polaron as a 
function of the frequency for different values of the charge carrier 
density at $T=0$. Three different structures appear in the normal 
state conductivity: 
a) a zero frequency delta function contribution; 
b) a strong band starting at $\omega=\omega_{0}$ that is the 
overlap of two components: a contribution from the intraband 
process and a  peak due to the 
polaron transition from the ground state to the first relaxed 
excited state; 
c) a smaller band at higher frequency due to the Frank-Condon transition of 
the polaron. 
The identification and characterization of these structures stem from
the detailed analysis of the polaron absorption in the case of non
interacting large polarons\cite{Devreese_review}.
Increasing the charge carrier density, we find that the large polaronic band 
due to the excitation involving the relaxed states ( b contribution) tends to 
move towards lower frequencies while its intensity decreases in favor of the 
rise of a Drude-like term around $\omega=0$ (see Fig.2a and Fig.2b). 

In particular, at $T=0$ the Drude weight D, i.e., the coefficient of the zero 
frequency delta function contribution,  
is determined making use of the sum-rule for 
the real part of the conductivity (see Eq.(\ref{11r})). This allows 
an estimate of the effective polaron mass $M_{eff}(n)$ as a function of 
the electron density, 
being the Drude weight D related to $M_{eff}(n)$ by the following expression: 
\begin {equation} 
D=\frac{\pi n e^2}{2 M_{eff}(n)}. 
\end {equation}

In table 1 the values of $M_{eff}(n)$ for $\alpha=5$ and 
$\alpha=6$ are reported. It is evident that, increasing the charge carrier 
density, the screening of the e-ph interaction increases and the polaron mass 
is reduced, tending to the band mass for large values of $n$. This behavior
is a confirmation of the trend obtained at weaker coupling in a model
that includes  polaron screening but neglects exchange 
effects\cite{iadonisi_1}.

In Fig3a we plot the normal state conductivity for larger 
electron densities at $T/T_{\scriptscriptstyle{D}}=.5$, where 
$T_{\scriptscriptstyle{D}}=\hbar\omega_0/k_{\scriptscriptstyle{B}}$ and 
$k_{\scriptscriptstyle{B}}$ is the Boltzman constant. 
Increasing the charge density, we find that the optical absorption, in 
agreement with experimental data in the metallic phase of the 
cuprates \cite{26,10,uchida}, is more and more controlled by the 
Drude-like term and, only for very high electron densities, no signature of 
the b) contribution is left . 
 
The behavior of the polaron absorption with 
temperature is also of interest. As shown in Fig.3b, with increasing $T$  
there is a transfer of 
spectral weight towards higher frequencies. Moreover, the intensity of the
contributions b) and c) increases with decreasing the temperature saturating
at $T/T_D\simeq 0.5$.

\section{Discussion}
The effects shown are very intriguing in connection of recent measurement
of infrared absorption in cuprates\cite{review calvani}.
In fact several experiments on the infrared response 
of the cuprates have 
pointed out that the optical absorption exhibits features which are 
common to many families of high-$T_c$ superconductors. 

In particular, the infrared normal state conductivity does not diminish with 
frequency as rapidly as one expects from a simple Drude picture\cite{2}. 
This behavior has been interpreted in terms of two different models: 
the anomalous Drude model\cite{3} and the multi-component or Drude-Lorentz 
model\cite{4}. 
In the former approach, the infrared conductivity has been 
attributed to a contribution from free charge carriers with an 
$\omega$-dependent scattering rate (arising, for example, by strong 
interactions with spin waves\cite{2}). 
In the Drude-Lorentz approach, instead, one assumes that the absorption is the result
of the superposition of different structures which  
can be identified in the conductivity spectra: 
1) a Drude-like peak centered at zero frequency; 
2) infrared-active vibrational modes (IRAV)\cite{6}; 
3) a broad excitation in the infrared band which is constituted by 
two different components: one temperature-independent around $.5\, 
eV$ ($\sim 4000\, cm^{-1}$)\cite{9,10} and the other strongly dependent on 
temperature with a peak around $.1\, eV$ ($\sim 1000\, cm^{-1}$)\cite{11,12} 
( d band); 4) the charge transfer band (CT) in the visible range which is 
attributed to the charge transfer transitions between $O_{2p}$ and 
$Cu_{3d}$ states. The Drude, the d band and IRAV contributions 
depend on the charge density injected by the doping and a significant
transfer of spectral weigth from the d band and Irav contributions to the
Drude peak is observed increasing the doping.
 To the d band and the IRAV, which appear when extra 
charge are injected into the lattice of a cuprate, it has been assigned a
polaronic origin both 
in electron-doped and hole doped compounds\cite{12,kim}. 
However, there is no general consensus on the type of polarons involved
in the absorption. In particular, the $1000\, cm^{-1}$ 
feature has been attributed to optical transitions 
involving small\cite{13}, large polarons\cite{emin,15} or both types of 
polarons\cite{eagles}. In any case, as first noted by 
Bi and Eklund\cite{25}in 
$La_{2-x}Sr_{x}CuO_{4+\delta}$, the polaron tend to be 
small in the dilute limit (small $x$) whereas in the limit of large $x$  
the polaron tend to expand, i.e., the polaron size increases with $x$ 
tracking a dopant-induced transformation to metallic conductivity. 

Finally we note that, recently, Calvani et al.\cite{27} have measured the infrared 
absorption of four 
different perovskite oxides observing an opposing behaviour between 
the cuprates and the non cuprates. In particular, in $La_2NiO_{4+y}$ 
and in $SrMnO_{4+y}$, in which there is a strong evidence of the presence 
of small polarons, the minimum of the d band deepens at low $T$ by a 
transfer of spectral weight towards higher energies. 
On the contrary, in the slightly and heavily doped cuprates at low 
temperatures the minimum of the d band tends to be filled by a transfer of 
spectral weight in the opposite direction (see also F. Li et al.\cite{li},) 
as it happens for a large polaron system previously described.  

The above scenario shows that the normal state conductivity of an 
interacting large polaron system presents features common  to the 
optical absorption of the cuprates\cite{note}. 
Even if the similarities found are not
conclusive in assigning to large polarons the main role in the absorption
of cuprates - more accurate calculations are nedeed to attempt a quantitative 
comparison between theory and experiments - we believe that the inclusion of 
long range Coulomb interactions among polarons is essential in understanding
whether infrared absorption in cuprates can be assigned to large polarons.

\section{Conclusions}
In this paper we calculated the normal state conductivity of 
a large polaron system including the electrostatic polaron-polaron interaction 
within the R.P.A. approximation. The approach 
recovers the optical absorption of non interacting Feynman polaron\cite{23} 
and allows to 
introduce in a perturbative way the many polaron effects. 
With increasing charge carrier density, 
we have found 
evidence of a transfer of spectral weight of the optical absorption   
towards lower frequencies: the Drude-like contribution increases as well as 
the screening effects of the e-ph interaction. Moreover, with increasing 
temperature, there is a transfer of spectral weight of the infrared absorption 
towards higher frequencies.  
Both these behaviours are observed in the infrared spectra of cuprates.

\section*{Acknowledgments}
We thank P. Calvani, S. Lupi and A. Paolone for helpful discussions, and
acknowledge the partial support from INFM.  
   
\section*{Table 1 captions}
\begin {description}
\item{Tab.1.} Effective polaron mass, in units of the electron band mass, as a 
function of the charge carrier density. The value of $n_0$ is  
$n=1.4\cdot10^{-5}$ in terms of $R_p^{-3}$, $R_p$ being the 
Fr\"ohlich polaron radius.
\end {description}

\section*{Figure captions}
\begin {description}
\item{Fig.1.} Diagrammatic representation of the Eq.(\ref{4r}). 
The solid line indicates the polaron propagator, the phonon is given by the 
dashed line, the dotted line describes the Coulomb interaction and the 
dotted-dashed line represents the incident photon.

\item{Fig.2.} a) Optical absorption per polaron, at $T=0$, as a function of the 
frequency  for different electron densities: 
$n=1.4\cdot10^{-5}$ (solid line), 
$n=1.4\cdot10^{-4}$ (dashed line), $n=1.4\cdot10^{-3}$ (dotted line), 
$n=1.4\cdot10^{-2}$ (dotted-dashed line); b) Optical absorption per polaron at 
finite temperature, $T/T_{\scriptscriptstyle{D}}=.5$, for 
different values of the charge carrier density: $n=1.4\cdot10^{-4}$ 
(dashed line), 
$n=1.4\cdot10^{-3}$ (dotted line), $n=1.4\cdot10^{-2}$ (dotted-dashed line). 
The value of $\epsilon_{0}/\epsilon_{\infty}$ is $3.4$. 
The electron density is measured in units of $R_p^{-3}$, $R_p$ being the 
Fr\"ohlich polaron radius, and the conductivity $\sigma(\omega)$ is expressed 
in terms of $ne^2/m\omega_0$. The value of $R_p^{-3}$ is $7\cdot10^{20}\, 
cm^{-3}$ when $\omega_0=30\, meV$ and $m=m_e$, 
where $m_e$ is the electron mass. 
   
\item{Fig.3.} a) Optical absorption per polaron at finite temperature, 
$T/T_{\scriptscriptstyle{D}}=.5$, for 
different values of the charge carrier density: $n=4.5\cdot10^{-2}$ 
(solid line), 
$n=1.4\cdot10^{-1}$ (dashed line), $n=.28$ (dotted line), 
$n=1.4$ (dotted-dashed line). The 
electron density is measured in units of $R_p^{-3}$, $R_p$ being the 
Fr\"ohlich polaron radius, and the temperature is given in 
units of $T_{\scriptscriptstyle{D}}=\hbar\omega_0/k_{\scriptscriptstyle{B}}$, 
$k_{\scriptscriptstyle{B}}$ being the Boltzman constant. 
In the inset is plotted the polaron mobility ($\mu$), i.e.  
$Re\sigma(\omega\rightarrow 0)$, as a function of the charge 
density for $\omega_0=30\, meV$, $\epsilon_{\infty}=3$ and $m=m_e$ where 
$m_e$ is the electron mass; b) Optical absorption of a single Feynman polaron 
for different values of $T$: $T/T_{\scriptscriptstyle{D}}=0$ (dashed line), 
$T/T_{\scriptscriptstyle{D}}=.5$ (dotted line), 
$T/T_{\scriptscriptstyle{D}}=1$ (dotted-dashed line). 
\end {description}

\newpage

\begin{center}
\begin{tabular}{|l|l|l|l|l|}
\hline
& n$_0$ & 10 n$_0$ & 10$^2$n$_0$ & 10$^3$n$_0$ \\ \hline
$\alpha $=5 & 3.56 & 3.07 & 2.19 & 1.42 \\ \hline
$\alpha $=6 & 6.21 & 5.28 & 3.58 & 1.94 \\ \hline
\end{tabular}

\bigskip 

Table 1
\end{center}

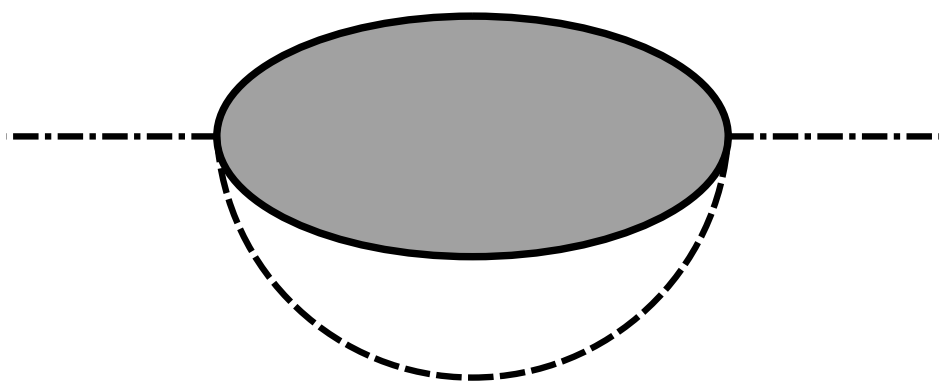
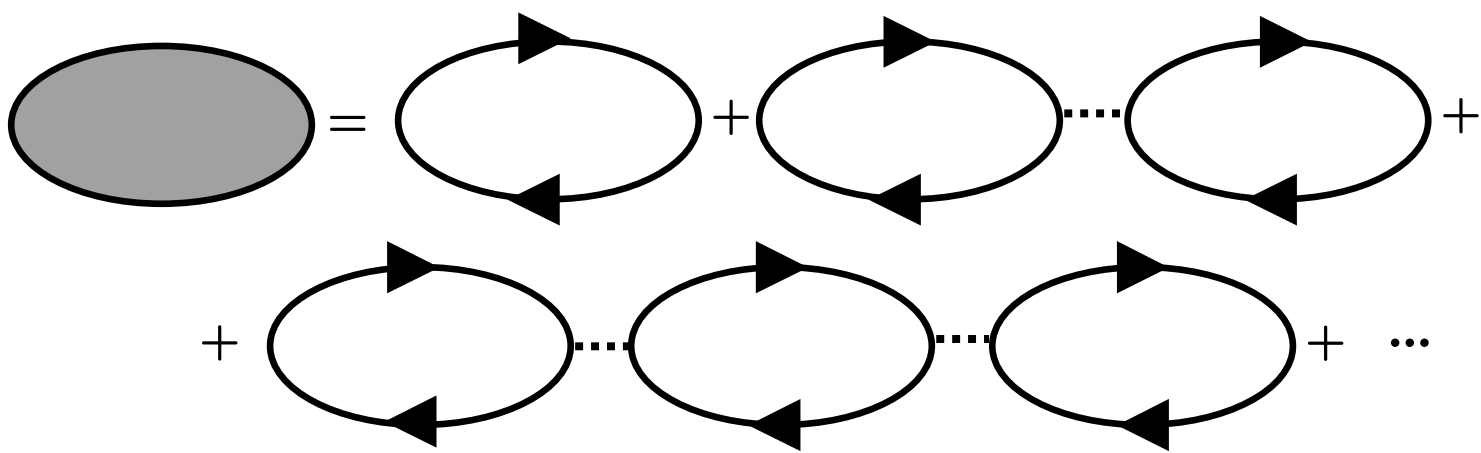
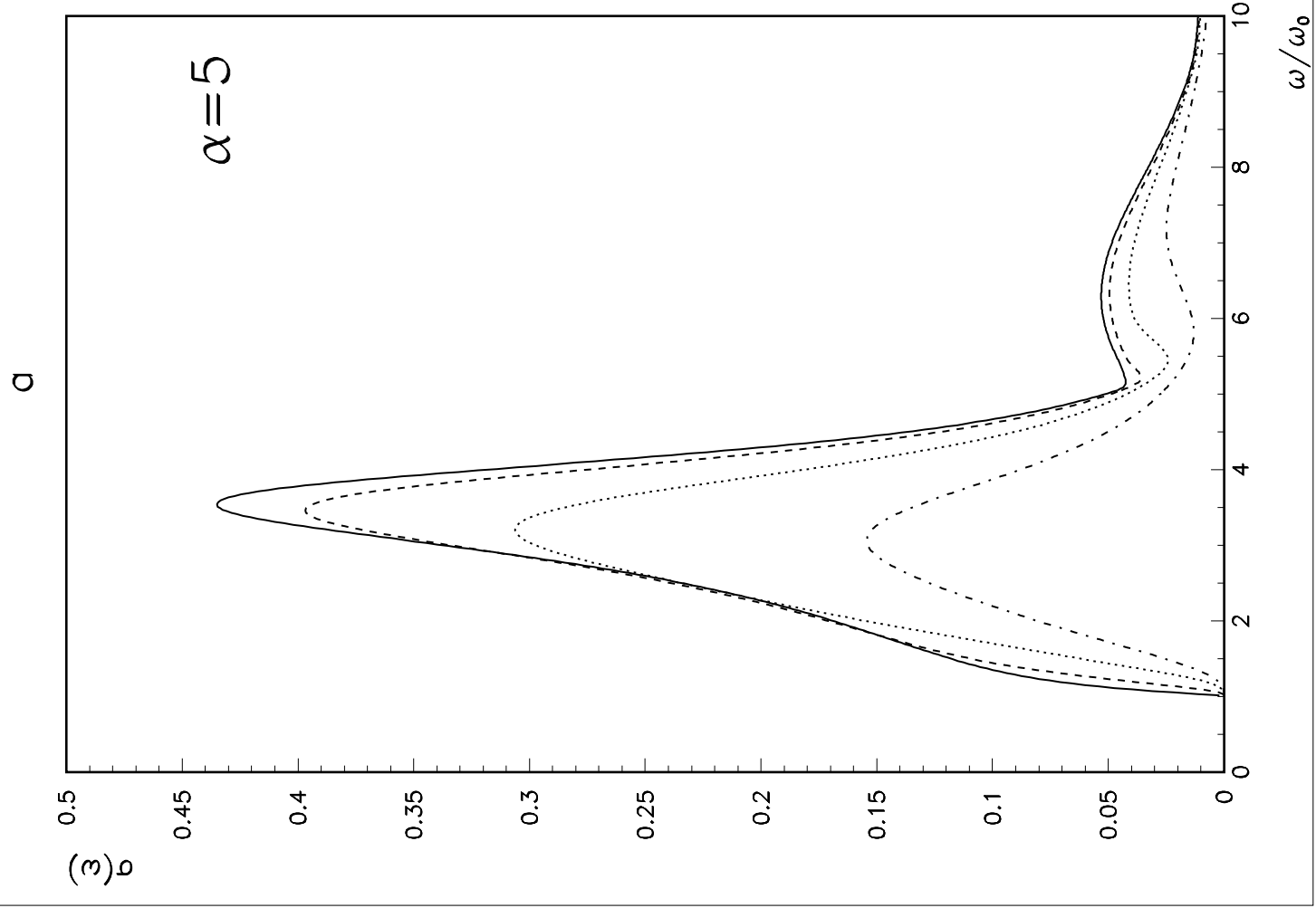
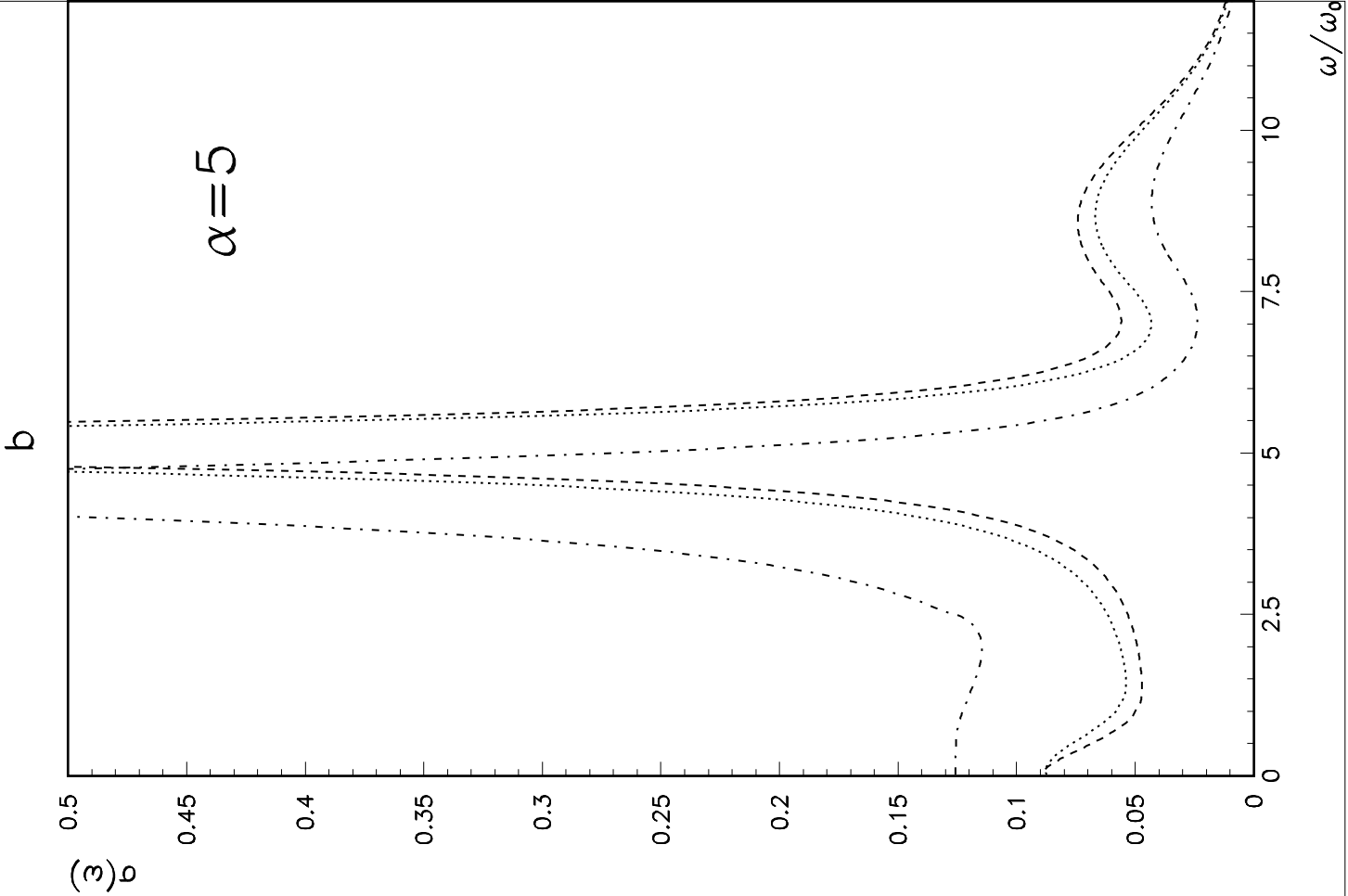
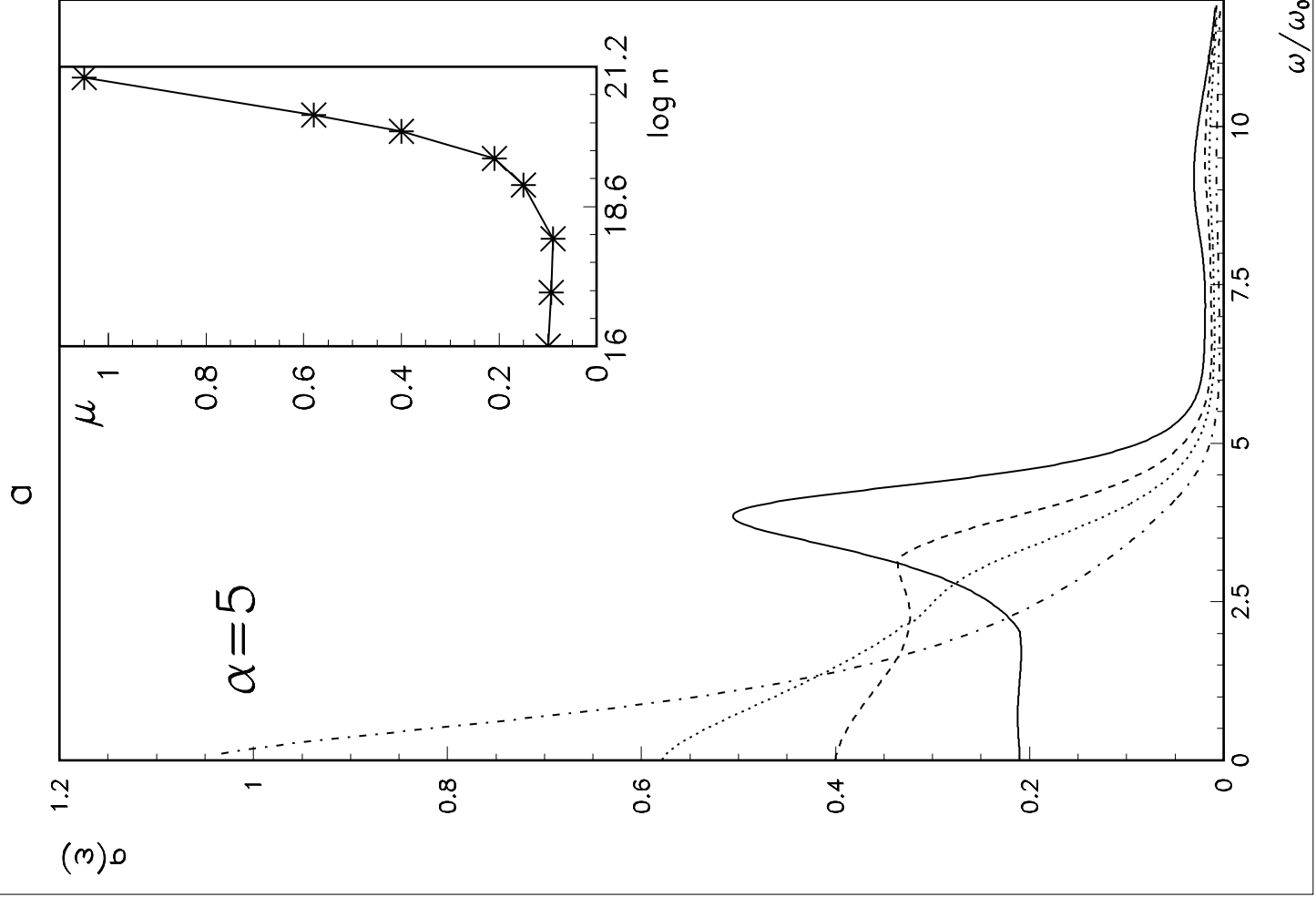
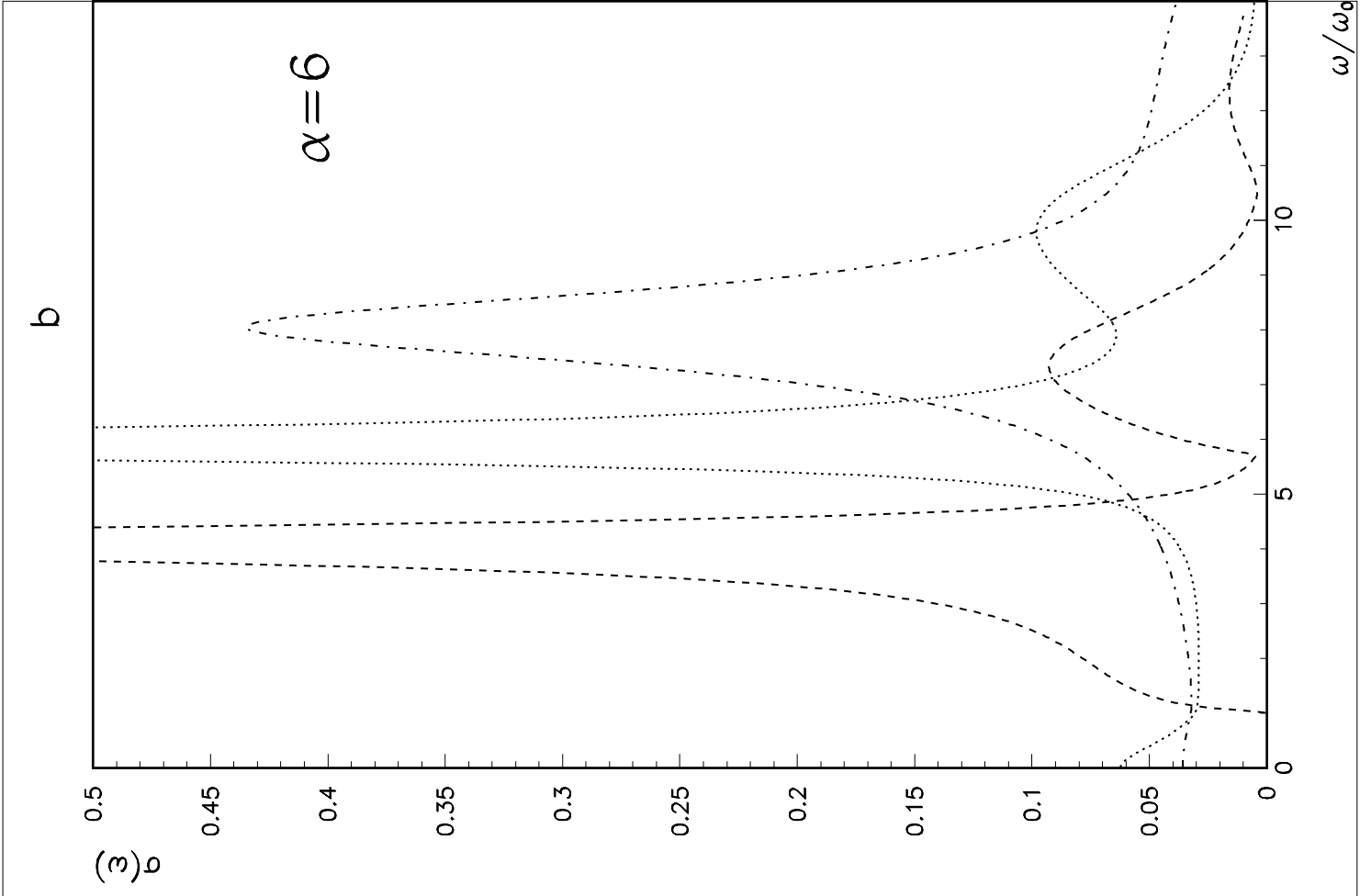
\end{document}